\documentclass[a4paper,12pt]{article}

\usepackage[english,russian]{babel}
\usepackage{latexsym} 
\usepackage{amsmath}
\usepackage{amssymb}
\usepackage{graphicx}
\usepackage{wrapfig}
\usepackage{sectsty}
\usepackage{indentfirst}

\allsectionsfont{\centering}
\makeatletter
\def\@seccntformat#1{\csname the#1\endcsname.\quad}
\makeatother

\newcommand{\be}{\begin{equation}}
\newcommand{\ee}{\end{equation}}
\newcommand{\ba}{\begin{eqnarray}}
\newcommand{\ea}{\end{eqnarray}}
\let\f\frac

\newcommand{\ds}{\displaystyle}
\def\S{Schwarzschild}

\begin{document}

\title{\Large\textbf{On Boundedness of the Admissible Time \\
Slowing
Down by the Gravitational Field}}

\date{}

\author{\normalsize S.\,S.
Gershtein\footnote{Semen.Gershtein@ihep.ru},
A.\,A. Logunov\footnote{Anatoly.Logunov@ihep.ru},
        M.\,A. Mestvirishvili,\\
{\small{\it Institute for High Energy Physics, Protvino, Russia}}}
\maketitle

\begin{abstract}
It is shown that there exists,  in the field theory of
gravitation, contrary to the General Theory of Relativity (GTR), a
{\it bound} for admissible time slowing down by the
gravitational
field which excludes a possibility of {\it unbounded compression} of
matter by the {\it gravity forces.}
\end{abstract}

In the Relativistic Theory of Gravitation (RTG) gravitational field
is considered as a physical field in the spirit of Faraday--Maxwell
evolving in the Minkowski space. In such an approach the conserved
energy-momentum tensor of matter and gravitational field is a source
of the field~[1,2], just as, in electrodynamics, the conserved
electric current is a source of electromagnetic field. Such a
universal source of gravitational field leads to an ``effective
Riemannian space'' of {\it simple topology}. Because the
gravitational field, as all other physical fields, evolves in the
Minkowski space, the fundamental conservation laws of
energy-momentum and angular momentum take place in the RTG, contrary
to the GTR. This means that special principle of relativity holds
rigorously for all physical fields including the gravitational one.

All these properties of the RTG distinguish it in essence from the
GTR and lead to a different system of gravitational equations. There
are, however, common features, e.g., that the gravitational field is
tensorial. In this paper we will show, taking as an example an
exact solution to the GTR equations, how physical consequences of
the RTG and GTR differ in the strong gra\-vi\-ta\-tio\-nal field.

In the GTR {\it the Hilbert--Einstein equations} for the spherically
symmetric static problem,  defined by the interval
\be
ds^2=c^2U(W)dt^2-V(W)dW^2 -W^2(d\theta^2+\sin^2\theta\,d\phi^2)\,,
\label{eq1} \ee
and the equation of state for a gas of relativistic
particles \be p(W)=\f{\,c^2\,}{3}\rho(W)\,, \label{eq2}
\ee
have an{\it 
exact solution} of the form
\be \rho(W)=\f{a}{W^2},\quad
a=\f{3}{7\varkappa},\quad \varkappa =\f{8\pi G}{c^2}\,, \label{eq3}
\ee
where $G$ is the gravitational constant.

From the matter equation,
\be
\f{\,1\,}{c^2}\f{dp}{dW}=-\Bigl(\rho+\f{p}{c^2}\Bigr)\f{1}{2U}\f{dU}{
dW}\,,
\label{eq4}
\ee
we find, making use of (2) and (3),
\be
\rho U^2=\alpha\,,
\label{eq5}
\ee
here  $\alpha$ is  an integration constant.

Substituting  (\ref{eq3}) into (\ref{eq5}) we obtain the expression
for the metric coefficient $U$ defining the time slowing down \be
U=\sqrt{\f{\,\alpha\,}{a}}\,W\,. \label{eq6} \ee
It is evident
thereof that when $W$ approaches the center the function  $U$
decreases, and an unbounded time slowing down occurs up to its stop
in the center, where  $U$ disappears. This means that there does not
exist, in the GTR, a bound for an admissible time slowing down by
the gravitational field. In this case the pressure, according to (2)
and (3), goes to infinity when approaching the center as
\be
\f{p}{c^2}
=\f{a}{3W^2}\,. \label{eq7} \ee
Since the pressure  $p$ is scalar,
this singularity at the center cannot be eliminated by the choice of
coordinates. This is also directly seen from the invariant
\[
R_{\mu\nu}R^{\mu\nu}=3\Bigl(\f{\,2\,}{7}\Bigr)^{\!2}\f{1}{W^4}\,,
\]
which also goes to infinity, when approaching the center.

Thus, if the matter inside a body obeys to the equation of state
(2), then, according to the GTR, the pressure and the density,
defined by expressions (\ref{eq3}) and (\ref{eq7}), become infinite
at the center of the body. It means that according to equations of
the GTR the forces of gravitational compression are un\-bo\-un\-de\-dly
strong. All this is a consequence of the {\it main cause}: the
absence in
the GTR of a bound for admissible time slowing down by gravitational
field. The absence of such a bound essentially contradicts to the
very essence of the GTR. For the given problem with density
(\ref{eq3}) the ball radius $R$  is related to the mass confined
inside the ball as follows
\[
R=\f{\,7\,}{3}W_g,\quad W_g=\f{2GM}{c^2}  \mbox{is the Schwarzschild
radius}.
\]
Nevertheless, though the ball radius exceeds, in this case, the
Schwarzschild
radius, the pressure and the density in the center of the ball,
achieve an infinite value.

In the Relativistic Theory of Gravitation the equations of the
gravitational field for the interval (1) have the form
\be
{\overset{\;\,\prime}{Z}}-\f{2Z}{U}{\overset{\;\,\prime}{U}}
-2\f{Z}{W}-\f{m^2W^3}{2}\Bigl(1-\f{U}{V}\acute{r}^2\Bigr)
=-\varkappa W^3\Bigl(\rho +\f{p}{c^2}\Bigr)U\,,
\label{eq8}
\ee
\be
1-\f{\,1\,}{2}\f{1}{UW}{\overset{\;\,\prime}{Z}}
+\f{m^2}{2}(W^2-r^2)
=\f{\,1\,}{2}\varkappa W^2\Bigl(\rho -\f{p}{c^2}\Bigr)\,,
\label{eq9}
\ee
\be
\mbox{where}\; Z=\f{UW^2}{V}\,.
\label{eq10}
\ee
Equations (\ref{eq8}) and (\ref{eq9}) under assumption that
\be
m^2(W^2-r^2)\ll 1,\quad \f{\,U\,}{V}\ll 1
\label{eq11}
\ee
can be somewhat simplified and, for the equation of state
(\ref{eq2}), take the form
\be
-U{\overset{\;\,\prime}{Z}}+2U^2W
=\f{\,2\,}{3}\varkappa \alpha W^3\,,
\label{eq12}
\ee
\be
U{\overset{\;\,\prime}{Z}}-2Z{\overset{\;\,\prime}{U}}
-\f{2ZU}{W}-\f{m^2}{2}UW^3
=-\f{\,4\,}{3}\varkappa \alpha W^3\,.
\label{eq13}
\ee
These equations do not change under the multiplicative
transformations
\be
U\to tU,\quad Z\to tZ,\quad m^2\to tm^2,\quad \alpha \to t^2\alpha\,.
\label{eq14}
\ee
With terms of the order of
$\beta W^2$ neglected equations
(\ref{eq12}) and (\ref{eq13}) have the solution
\be
U=\beta +\sqrt{\f{\,\alpha\,}{a}}\,W\,,
\label{eq15}
\ee
\be
Z=W^2\Bigl(\beta +\f{\,4\,}{7}\sqrt{\f{\,\alpha\,}{a}}\,W\Bigr)\,.
\label{eq16}
\ee
The constant  $\beta$ is non-zero and, due to multiplicative
transformations
(\ref{eq14}),  {\it is proportional to the gravitation rest mass
squared}.

That is the quantity   $\beta$ that defines the bound for the
admissible slowing down of time by a gravitational field. In the GTR
this quantity is equal to zero, and this fact leads to an unbounded
slowing down of time up to its stop, as well as to the infinite
values of density and pressure, and even to such non-physical objects
as
``black holes''. According to the RTG such mystic objects are absent
from Nature.

Comparing  (\ref{eq10}) and  (\ref{eq16}) we find for the second
metric coefficient
$V$ the expression
\be
V=\f{\beta +\sqrt{\ds\f{\,\alpha\,}{a}}\,W}
{\beta +\ds\f{\,4\,}{7}\sqrt{\ds\f{\,\alpha\,}{a}}\,W}\,.
\label{eq17}
\ee
Taking into accounts (\ref{eq15}) and (\ref{eq17}) we find
\[
\f{\,U\,}{V}=\beta +\f{\,4\,}{7}\sqrt{\f{\,\alpha\,}{a}}\,W\,.
\]
Due to smallness of the gravity $\beta$, which is proportional to
$m^2$, this expression for small values of $W$ is small enough in
compare with unity and this justifies approximation (11).

Making use of (15) in (5) we find for the density of matter,
$\rho$, the expression
\be
\rho =\f{\alpha}{\Bigl(\beta +\sqrt{\ds\f{\,\alpha\,}{a}}\,W\Bigr)^2}
\label{eq18}
\ee
It is evident thereof that the density of matter,
$\rho$, defined by expression (18), and the pressure of matter,
according to formula (2) are bounded due to the bound of the
admissible slowing down of time, $\beta$. So, in contrast with the
GTR, the forces of gravitational compression are always finite in the
RTG. It means that the gravitational field, slowing down the run of
time, cannot stop it, what is quite natural from physical standpoint.
It follows from the above analysis, as well as from the analysis of
the exact internal Schwarzschild solution made in [8], that physical
gravitational field in the RGT has a property to stop the process of
gravitational compression due to the bound of the admissible time
slowing down by gravitational field.
We call such a phenomena that follows from a physical property of
gravitational field to {\it restrict itself}  as {\it ``gravitational
selfstop''}.
So, an {\it unbounded} gravitational compression
of matter takes place
in the GTR; 
gravitational compression  of matter in the RTG 
{\it always bounded} 
and the {\it gravitational
selfstop} takes place. 

This makes the difference between the physical consequences of the
GTR and the RTG. Such a difference leads to essential changes both
in the Universe evolution [4] and in the process of collapse [5].

In conclusion the authors thank V.A. Petrov and N.E. Tyurin for
valuable discussions.

\renewcommand{\refname}
\end{document}